# Interarm islands in the Milky Way – the one near the Cygnus spiral arm

Jacques P Vallée

National Research Council of Canada, Herzberg Astronomy & Astrophysics Research Center, 5071 West Saanich Road, Victoria, B.C., Canada V9E 2E7   jacques.p.vallee@gmail.com

**Abstract.** This study extends to the structure of the Galaxy. Our main goal is to focus on the first spiral arm beyond the Perseus arm, often called the Cygnus arm or the 'Outer Norma' arm, by appraising the distributions of the masers near the Cygnus arm. The method is to employ masers whose trigonometric distances were measured with accuracy. The maser data come from published literature – see column 8 in Table 1 here, having been obtained via the existing networks ( US VLBA , the Japanese VERA, the European VLBI, and the Australian LBA). The new results for Cygnus are split in two groups: those located near a recent CO-fitted global model spiral arm, and those congregating within an 'interarm island' located halfway between the Perseus arm and the Cygnus arm. Next, we compare this island with other similar interarm objects near other spiral arms. Thus we delineate an interarm island (6x2 kpc) located between the two long spiral arms (Cygnus and Perseus arms); this is reminiscent of the small 'Local Orion arm' (4x2 kpc) found earlier between the Perseus and Sagittarius arms, and of the old 'Loop' (2x0.5 kpc) found earlier between the Sagittarius and Scutum arms. Various arm models are compared, based on observational data (masers, HII regions, HI gas, young stars, CO 1-0 gas).

## 1. Introduction.

The advance of our knowledge of the Milky Way disk comes slowly, with new observational data coming up from time to time. Recent data have begun to map the Cygnus spiral arm, in Galactic Quadrants II and III, about 15 kpc from the Galactic Center (about 7 kpc from the Sun) – see Quiroga-Nunez et al (2019) and Reid et al (2019).

The discovery of radio masers located inside high mass star forming regions allows precise distance estimates to be made through the parallax method. The astrometric surveys, through very long baseline interferometry, notably the VERA (Sakai et al 2015) and BeSSeL (Reid et al 2014) projects, have produced quite a few precise distance estimates.

The precise distance of masers is necessary to compare the location of a maser to the location of other spiral arm tracers such as gas and dust. Some discrepancies in the trigonometric distances to a maser have been noted. Toward the Cygnus spiral arm, the maser in S269 (G196.45-01.48) had been measured trigonometrically before, with different values: Honma et al (2007) found a distance of 5.28 ± 0.23 kpc, while Asaki et al (2014) found a distance of 4.05 ±0.60 kpc, and Quiroga-Nunez et al (2019) found 4.15 ±0.22 kpc and explained the distance discrepancy to fitting a 'problematic morphology' versus fitting a 'compact core'.

With partial data come partial maps, and perhaps partial explanations. For example, the map of Quiroga-Nunez et al (2019 – their Fig. 7) and of Reid et al (2019 – their Fig. 2) contains a possible logarithmic spiral arm for Cygnus, which they choose to misalign with a 'kink' at galactic longitude 140$^o$, with 'segment 1' (below l=140$^o$, pitch angle of 10.5$^o$ ) and 'segment 2' (after l=140$^o$, pitch angle of 7.9$^o$). They claim that the 'reconstruction' of the Cygnus arm is still the best procedure with limited data, using short segmented arms. Yet no physical explanation (galactic tides, inbound halo streams, etc) is provided for introducing this kink. We are motivated by the question: is there another alternative to such a sudden kink?

Several nearby spiral galaxies show a large global pattern, such as long arms around the galactic nucleus, as well as some luminous matter grouped into interarm islands, appearing here and there between the long arms.

In the literature, numerous terms have been used to describe local agglomeration outside the long arms – here we will call the bigger ones as 'interarm islands' instead of: armlets, blobs, branches, bridges, deviations, feathers, features, fingers, forks, loops, rings, spurs, swaths – see an example in Figure 3 in Vallée (2018b).

How are interarm islands formed? Independently or a derivative of existing arms? It is natural to search for independent theoretical models to build the interarm islands, requiring that the islands are not massive enough to disrupt the global arm model. It is also tempting to want to 'attach' an island to a nearby arm, in a tenuous way, which necessitates to insert some new localized physical functions into the existing global arm models. Dwarf galaxies? The fact that our Galaxy has 'eaten up' other nearby small galaxies in the past points to localized disruptions when ingesting portions of a galaxy at each orbital encounter. The ingesting of various portions of the trail of the nearby Sagittarius dwarf galaxy over a long period of time has been mentioned before - near the Cygnus arm, near the Perseus arm (Olano 2016; Laporte et al 2018) and near the Orion ring or 'Local arm' (Vallée, 2018b). Resonance ? For interarm islands to be included in a larger-scale theoretical model, one looks for a 'banana-like' shape function as predicted near the co-rotation radius (Lépine et al 2017), or else for a signature of the 4:1 Outer Linblad Resonance from a slow and long galactic nucleus bar (Hunt and Bovy 2018), or else for a signature of a 8:1 Inner Linblad Resonance (Mitchenko et al 2018). However, one does not expect a strong resonance in between each long spiral arm. Nuclear bar ? The effect of a bar located in the galactic nucleus is possible, but at the distance of the Sun (8.1 kpc from the Galactic Center) and beyond, the resonance model of Lépine et al (2017 – their Section 3) with a realistic bar strength showed negligible effects and did not distort the long spiral arms. Numerical models with large N-body simulations have shown non-stationary self-excited spiral dynamic patterns, forming multiple arm segments (Baba 2015).

      Here our motivation is to re-examine the nearby interarms and their distances to the nearest global spiral arms (Section 2). Our new results show that the situation near the Cygnus arm can be explained by the presence of a new interarm island located between the two long spiral arms (Cygnus and Perseus). Section 3 presents our concept of an interarm island below the Cygnus arm and above the Perseus arm. Section 4 deals with the observed maser offsets from the Cygnus arm. The discussion in Section 5 deals with other arm models, notably the difficult reconstruction of segmented arms together, the use of optical Gaia DR2 parallaxes from faint stars, HII arms, HI arms, and optical arms. We conclude in Section 6.

## 2. Trigonometric masers and a new 4-arm model

      **Table 1** shows the trigonometric masers near the Cygnus spiral arm. In this Table 1, the columns show the source name, galactic coordinates, distance (=1/parallax), distance range [1/(parallax + error); 1/(parallax – error)] distance

error (half of the range), f value (fractional parallax error), systemic velocity and errors, and reference to these data. All of the maser data here have a fractional parallax error f of 20% or less, so do not require a bias correction (Bailer-Jones 2015).

The maser G211.60+01.06 in Table 4 of Quiroga-Nunez et al (2019) has been assigned previously to the Perseus arm (Reid et al 2014), owing to its close proximity. The maser G073.65+00.19 in Table 4 of Quiroga-Nunez et al (2019) had a fractional parallax error > 20% and is omitted here.

Global arm model. A global spiral arm model must take advantage of a large angular view, fitting arm tangents in two galactic quadrants (I and IV). The two tangents from the Sun to a spiral arm, seen in Galactic Quadrant at galactic longitude $l_I$ and in Galactic Quadrant IV at $l_{IV}$, can give the pitch angle p through the equation:

$$\ln[\sin(l_I)/\sin(2\pi - l_{IV})] = (l_I - l_{IV} + \pi) \cdot \tan[p]$$

The use of two Galactic quadrants provides a more precise global arm pitch angle value, revealing the Milky Way's global order and its mirror-image symmetry as one crosses the Galactic Meridian. Thus, for the Sagittarius arm, the observed arm tangents at l=50.5° ±0.5° and at l=283° ±1° yielded a mean pitch angle near -13.9° ±0.6° over various arm tracers (Vallée 2015 – his table 1). Similar pitch data for the Scutum arm at l=32° ±1° and at l=310° ±1° yielded a mean pitch angle near -13.3° ±0.5° over various arm tracers (Vallée 2015 – his Table 2). The Norma arm, at l=18° ±2° and at l=329° ±1° yielded a mean pitch angle near -13.7° ±1.4° over various arm tracers (Vallée 2017b – his Table 1). The 4-arm model is explained further in Section 7 and figure 5 in Vallée (2017c). A longer review of the arms in the Milky Way was provided (Vallée 2017d).

When scanning of a radio telescope in galactic longitudes, the CO 1-0 gas shows a greater intensity at a specific galactic longitude for each spiral arm, when the CO is observed tangent to a spiral arm. The model arm tangents in galactic longitudes, as seen from the Sun, were fitted to the published observed CO 1-0 gas intensity peak (as catalogued in Table 5 of Vallée 2016a), using a telescope beam of 8.8' scanning in galactic longitudes.

The 4-arm model chosen here is that of Vallée (2017b and 2017c), fitting the galactic longitudes of the peak intensity in the diffuse CO 1-0 gas, using a Sun's distance to the Galactic Center of 8.0 kpc, the origins of the four arms at 2.2 kpc from the Galactic Centre (Vallée 2016b), a pitch angle at -13.1°, and an orbital circular velocity at 230 km/s (Vallée 2017a).

Here we tweaked this CO-based model, to incorporate the latest value for the distance of the Sun to the Galactic Center (8.12 using S2 precession - Abuter et al 2018a; 8.18 kpc using the S2 star - Abuter at al 2018b), averaging 8.15 kpc. We also incorporated the latest value for the circular orbital velocity of the local standard of rest made of the stars near the Sun (233.3 km/s - McGaugh 2018; 233.4 km/s – Drimmel & Poggio 2018), averaging 233.35 k/s. We did not employed masers to obtain a model fit to the spiral arms.

## 3. New and old interarm islands

### 3.1 Choice of an arm model and of tracers

In this paper, the locations of the observed trigonometric radio masers are compared to the more precise location of the CO 1-0 based model arm as obtained from the global view using two galactic quadrants (see above). Other arm models are compared elsewhere (Table 3 in Vallée 2017c) or discussed later (Section 5). Not all observed masers a relocated in spiral arms – some are located in interarms.

**Table 2** compiles what is currently known about these interarm islands. In this table, column 1 gives the interarm island name, while the two nearest long spiral arms are given in column 2. Its galactic longitude (column 3) and radial velocity (column 4) follow. Next comes its distance (column 5), and the interarm island width (column 6). A reference follows in column 7.

**Figure 1** shows in red the distribution of masers (filled squares; source data in column 8 of Table 1) and the arm middle (CO-fitted model curve) for the Cygnus (Outer Norma) arm. Similar data (masers, model) for the Perseus arm are shown in yellow, with data from Vallée (2018a). Also, similar data (masers, model) for the Sagittarius arm are in green, with data from Vallée (2019).

In addition to the global CO-based 4-arm model, we plotted some small observational features in between some long arms.

-Thus the interarm island (ring) around the Sun (orange contour) encompasses local stars near the Sun, as studied by Vallée (2018b – his Fig.1 and Tables 1 and 2). Numerous tracers (stars, masers, dust, HI, etc), each with a different distance estimate and its error, have been used to map the 'Local Arm' (or interarm island, near the Sun) yielding a complex interarm island shape (Vallée 2018b).

-The interarm island, in between the Scutum and the Sagittarius arm (orange contour), near galactic longitudes from l=32° to 39°, was discussed in Rigby et al (2016 – their Section 3.4 and Fig. 6) and called a 'potential loop', being around a radial velocity from +605 to +90 km/s. The interarm island's kinematical distance was obtained through the equations 1 and 2 in Roman-Duval et al (2009), for the 'near' distance along the line-of-sight (l=32°, $v_r$=+75 km/s, d= 4.2 kpc; l=39°, $v_r$=+75 km/s, d= 4.6 kpc).

-As proposed here, the interarm island, in between the Cygnus arm and the Perseus arm (orange contour), encompasses several masers (in red), but it does not encompass the masers closer to the Cygnus arm model (e.g., G182.67, G135.27, G097.53). The separation of masers (at the inner edge of a spiral arm) and the CO-based arm model should be about 600 pc or less (see Section 4 below), and thus masers farther from a spiral arm (defined by dust lane or CO gas) will be in an interarm. Along the Galactic Meridian the Cygnus arm near 15.7 kpc and the Perseus arm near 10.9 kpc give an interarm of 4.8 kpc, and masers further away inward from the Cygnus arm by 0.6 kpc or more are located in the interarm.

This proposed interarm island between the Cygnus and Perseus arms thus resembles somewhat the observed one around the Sun, between the Perseus arm and the Sagittarius arm. A good illustration of the objects in the interarm around the Sun was shown in Figure 1 and Table 1 in Vallée (2018b), including some masers in a 'spur', some HII regions in a 'branch', some CO gas in a 'layer', some open star clusters in a 'bridge' or a 'ring', and some Cepheids in a 'group'.

Elsewhere, to avoid interarm islands, it was proposed to re-orient the Cygnus arm at a greater galactic longitude than l=140°, causing this long arm to come closer to the Sun and to become wider in radial distance. Their alleged 'kink' at galactic longitude l=140° is shown (black dotted line) – see Quiroga-Nunez et al 2019 – their fig.7 and Section 4.4.2. The falling Cygnus arm at l> 140° would not explain the more distant masers there.

### 3.2  Interarm islands and spiral arm kinematics

**Figure 2** shows the 4-arm model in radial velocity versus galactic longitude, in Galactic Quadrants II and III. The observed masers (red squares) near the Cygnus arm (red curve) are plotted (from Table 1), in radial velocity (km/s) versus galactic longitude (°).

The interarm island around the Sun is seen at low radial velocities (orange contour), as proposed earlier (see Vallée 2018b – his Fig. 2). Also shown is the proposed interarm island between the Cygnus arm and the Perseus arm (orange contour).

**Figure 3** shows the CO-based 4-arm model in radial velocity and galactic longitude, in Galactic Quadrant I. Also shown is the interarm island between the Sagittarius and the Scutum arms (orange contour).

### 3.3 Legitimacy of a short interarm island, between Cygnus and Perseus

In particular, the Cygnus-Perseus interarm island, as discussed in Figure 1, has been seen very recently, including Cepheid variable stars and young open star clusters.

Thus, Molina-Lera et al (2019 – their Fig. 17) studied young open star clusters in Galactic Quadrant II, pointing out that several of them could be located in between the Perseus arm and the Cygnus (outer) arm. They noted that the young open star cluster Waterloo 1 could be in an 'interarm feature', being halfway between the Perseus and the Cygnus arms near longitude l= $150^o$ and solar distance near 4 kpc (their Section 5 and Figure 1). That fits near the middle of our interarm island in Galactic Quadrant II (in Fig. 1). They also showed ten young HII regions between l=$150^o$ and $180^o$ at left of Waterloo 1, closer to the Galactic Meridian but also located in the interarm between the Perseus arm and the Cygnus arm, fitting well with our sketch in Fig. 1 here.

**Figure 4** shows the young open star clusters Waterloo 1 (the asterisk near l=$150^o$, and near 4 kpc away from the Sun), from Molina Lera et al (2019). Also shown are the HII regions (open circles) in the interarm between Perseus and Cygnus arms, and l=$110^o$ to $210^o$ from Molina Lera et al (2019 – their Figure 17). It can be seen that the interarm island proposed here (orange contour) between the Perseus arm and the Cygnus arm, encompasses a suitable number of HII regions (open circles), as well as being very close to the young open star cluster Waterloo I (black asterisk), as well as Sharpless 207 from Yasui et al (2016a) shown here (red asterisk), and Sharpless 208 from Yasui et al (2016b) shown here (blue asterisk).

Also, mapping of the Milky Way disk with Classical Cepheid variable stars was done recently by Skowron et al (2019 – their Fig.1b), showing many Cepheids in between the Perseus arm and the Cygnus arm – see their interarm in Galactic Quadrant II (some Cepheids) and Quadrant III (twice more Cepheids) near the Galactic Meridian. These interarm Cepheids overlap well, in quantity ratio and in angular size, the interarm island sketched in Figure 1 here, in Galactic Quadrant III.

Galactic molecular clouds (GMC) are difficult to locate along the Galactic Meridian, as their distance estimates are mostly kinematical, and their radial velocities should be close to zero – the non-kinematical part would give a wrong value.

In general, these islands could possibly be due to local turbulent events (expanding supernovae,  expanding shells of stellar clusters, etc). The local temporary turbulence may co-exist with the large-scale density-waves, and may not erase the pattern of long spiral arms.     Inbound halo superclouds, and incoming dwarf spheroidal galaxies, as suitably segmented by tidal interactions with the Milky Way, may deposit some segments in a spiral-arm galaxy,  one segment at a time (Section 6.4 in Vallée 2018b; Fig. 5 in Olano 2016; Fig. 1 in Law et al 2005). Those localized segments as deposited in the disc galaxy over time may *co-exist* with the large-scale density-waves, and may not erase the pattern of long spiral arms (Antoja et al 2018; Ibata et al 1994).

All these stellar segments may be more conspicuous if located in the interarm ("islands"), rather than in an existing long spiral arm full of  stars with a different origin. Stars at a given location in the disk may come from distinct origins, each having a small difference in velocity or origin or chemical composition. Hence a stellar trail may be easier to detect in the interarm where there are few or no other stars (not originating from the trail itself). In the local interarm, we see different trails at different locations around the Sun's position in the Local Armlet (Fig. 2 in Hunt et al 2019).

Isolated star formation between arms is frequently seen. Extended optical maps of some nearby galaxies already show some localized interarm material ('islands') located in between the long spiral arms. Their appearances in nearby galaxies show that there are readily created over galactic times, and that the global density wave pattern is not readily erased by these temporary and localized events.

### 3.4 Legitimacy of long spiral arms, in the outer Galaxy

The legitimacy of the observational existence of long arms is already proven in many nearby spiral galaxies that are suitably isolated.  How far can the spiral arms carry on, with roughly the same spiral pitch angle ?    Where will this model of long logarithmic spiral arms with roughly the same pitch angle stop – out to which galactic radius ? Radio observations near a wavelength of 21 cm does allow the mapping of spiral arms beyond the same arms as seen at optical wavelengths.

**Table 3** assembles some disc galaxies (col. 1) at a  nearby distance  (col. 2) showing a spiral arm shape without a significant kink in pitch angle out to a galactic radius in arcmin (col. 3) and in kpc (col. 4), near a wavelength of 21cm in HI line or synchrotron continuum. Only some galaxies with an appreciable galactic radius longer than 12 kpc are listed here. It can be seen that for suitably isolated disc galaxies their arms can remain spiral for as far as 26 kpc, as long as the galaxy is sufficiently isolated in space.

The Cygnus arm is seen in Figure 1 near 15 kpc along the Galactic Meridian. It follows from Table 3 that the Cygnus (outer Norma) arm is sufficiently close to the Galactic Center and sufficiently far from the massive Large and Small Magellanic Clouds in order to retain its shape and pitch angle.

Also, the Cygnus arm is already determined by multiple arm tracers, each tracer with its own long 3-kpc bars in the top of Figure 1, along the Galactic Meridian: (i) the broad CO model, (ii) the HII regions model of Hou & Han (2014), and (iii) the dense HI gas model of Koo et al (2017).

More generally, going beyond the Sun along the Galactic Meridian, the Cygnus arm at a galactic radius of 15 kpc, the outer Scutum arm at 23 kpc, and the outer Sagittarius arm at 32 kpc are likely to retain their logarithmic shape (see Fig. 1a in Vallée 2017d), being less than about two-thirds of the distance to the Magellanic Clouds at 60 kpc; the shape of the outer Perseus arm at 44 kpc may be tidally affected by the Magellanic Clouds.

Going beyond the Galactic Center, along the Galactic Meridian, the inner Scutum arm at a galactic radius of -11 kpc, the inner Sagittarius arm at -15 kpc, inner Perseus arm at -22kpc, and the inner Norma arm -32 kpc are likely to retain their logarithmic shape (see Fig. 1b in Vallée 2017d), for the same tidal reason with the Magellanic Clouds.

Elsewhere, it was shown that the pitch angle of a logarithmic spiral arm stays roughly the same out to a large distance (Fig. 1 in Vallée 2016b), around a mean value (along the same arm, the local pitch may deviate slightly up or down from the large-scale global pitch).

## 4. The masers associated with the Cygnus arm are offset from the CO-based arm

For the large-scale spiral arm model employed here, as fitted to the CO 1-0 gas, then the presence of some trigonometric masers just below the Cygnus arm (in Fig. 1), but not part of the interarm island there, would constitute an inward offset from the CO 1-0 gas (long arm).

Three trigonometric masers in Table 1 are close to the Cygnus arm, and all three masers are near the inside arm edge (closer to the Galactic Center) – and not beyond the Cygnus arm:

-G182.67-03.26 is separated from the Cygnus CO-based model by 0.6 kpc;
-G135.27+02.79 is separated from the  CO-based arm model by 1.2 kpc;
-G097.53+03.18 is separated from the CO-based arm model by 1.3 kpc.

This gives a mean separation (offset) of 1.0 ± 0.4 kpc, inward from the CO-based arm.

This offset implies that the Cygnus arm (near a galactic radius of 15 kpc) is closer to the Galactic Center than the 'co-rotation' radius (where the arm pattern speed and gas speed are equal).  Using a circular orbital gas velocity of 233.35 km/s (see also Vallée, 2017a), the angular speed of the gas  $\Omega_{gas}$ = 15.6 km/s/kpc there, while the angular speed of the density-wave pattern must be slower.

In the density-wave theory, Roberts  (1975 – his Fig.2) predicts an offset between shocked dust (and masers) and the 'potential minimum' (broadly distributed  low-density CO gas peak intensity) of 3.7% of an arm cycle amounting to 465 pc when using 4 arms, at a galactic radius of 8 kpc. Dobbs and Pringle (2010 – their Fig. 4a) predict an offset of  2.8 %  of an arm cycle giving 352 pc for the same conditions. Gittins and Clarke (2004 – their Fig. 11) predict an offset of 9.6% of an interarm, giving 120 pc at a galactic radius of 8 kpc and m=4 arms, with a 1.2 kpc interarm.

In the Dynamic spiral arm theories, there is no flow across an arm, hence no age gradient nor offset of masers to older stars (Dobbs & Baba 2014). In one numerical model, the gas flow is tangent to an arm, not crossing it and not creating an age gradient on both sides of the arm – see Fig. 5 in Baba et al (2016).

In tidally-based theories of spiral arm formation, and theories where arms formed from a bar in the galactic nucleus,  two long arms are created, not four arms as seen in the Milky Way (Dobbs & Baba, 2014).

## 5. Discussion on various arm models

Our global 4-arm model employed CO tracers in both inner Galactic Quadrants I and IV, yielding a precise mean arm pitch angle over a wide angular view – see Section 2 above  on why to use our global arm model.  Other arm models have been proposed - here we discuss various problems associated with other arm models as found in the literature.

## 5.1 The segmented Reid et al arm model, constructed piecewises by adding short lines between the locations of radio masers

The segmented spiral arm model of Reid et al (2016 – their Fig. 1) and Reid (2017) has been built up to cover half of the Milky Way (mostly in Galactic Quadrants I and II). Doing that all over the Milky Way disk would in principle reconstruct the 4 long main spiral arms.

In practice, some masers in the interarms could be misconstrued as masers in a long arm, thus altering the pitch angle of that arm. The inherent incompleteness of maser data would not allow one to easily separate masers in interarm islands (or bridges) from masers in long arms.

For examples, the use of trigonometric masers near the Cygnus arm has produced pitch angle values of -6.2° (Quiroga-Nunez et al 2019), -14.9° (Hachisuka et al. 2015), -13.8° (Reid et al 2014), -11.6° (Sakai et al 2012), and -2.3° (Reid et al 2009), among others. The published pitch angle error values (near 2°) are small when compared to the wide range of the pitch angle values (from -2° to -15°). For the Cygnus arm, the pitch angle was often computed with the early problematic S269 results (see Section 1), except for the later results of Hachisuka et al (2015) and Quiroga-Nunez et al (2019).

Similarly, the use of trigonometric masers near the Perseus arm has yielded pitch angle values from -9.0° (Zhang et al 2019), to -13.0° (Reid 2012), up to -17.8° (Sakai et al 2012). Again here, the pitch angle range (-9° to -18°) is wider than the pitch angle error value (near 2°). Claims have been made that some long spiral arms may not have masers in some segment, notably the Perseus arm between galactic longitude 50° to 80° (Zhang et al 2013). A full list of published pitch angle for each arm can be seen in Vallée (2017c).

It is thus difficult to extrapolate an arm segment with a claimed pitch angle value (see above), when that pitch angle value is likely not obtained for the whole global arm over two Galactic Quadrants.

Their Cygnus arm model (Reid et al 2016 – their Fig.1), fitted to radio masers near that arm and in the interarm, is at 4.9 kpc beyond the Sun, using their value of $R_{sun}$=8.34 kpc (see top of Figure 1 here).

*Kinks galore ?* An updated model by Reid et al (2019 – their Fig. 2) fitted new masers to their old spiral arm model; they introduced several 'kinks' between segments of spiral arms, with an inner kink near the beginning of the Cygnus arm near the Galactic Center. They fitted the Cygnus arm through all the masers in the interarm islands, with an outer kink following Quiroga-Nunez et al (2019).

They 'constrained' the galactic longitude tangent of the Scutum-Centaurus arm at l=306.1° (their Table 2), yet all observed other arm tracers are *outside*, namely - the broad CO 1-0 gas at l=309° and of the dust 240μm and 60μm at l=311° (table 3 in Vallée 2016a); such a dust-masers offset of 5° at a solar distance of 6 kpc corresponds to a deviation of 524 pc (unexplained).

Ditto for the Norma arm tangent which they 'constrained' at l= 327.5° (their table 2), yet all observed tracers are *outside*, namely between 328.4° (CO between 328° and 330° – table 5 in Vallée 2016a) and 332° (dust 240μm and 2.4μm between – Table 3 in Vallée 2016a); such a dust-masers offset of 4.5° at a solar distance of 7 kpc corresponds to a deviation of 550 pc (unexplained).

Ditto for the Sagittarius-Carina arm tangent which they 'constrained' at l= 285.6° (their table 2), yet all observed tracers are *outside*, namely between 285° (dust 60 μm – table 3 in Vallée 2016a) and 281.3° (CO between 280° and 283° – table 5 in Vallée 2016a).

Their '3-kpc-arm' arm has a round shape (yellow ring in their fig. 1), with arm tangents at l=337° and at l= 28°. Their '3-kpc-North arm' at l=337.0° (their table 2) is shown elsewhere as the 'Start of the Perseus arm' at the same longitude (see table 3 in Vallée 2016a). Their '3-kpc arm' at l=28° is shown elsewhere as part of the start of the Scutum arm (see table 7 in Vallée 2016a). Their use of the two 3-kpc arms as a single circular ring does not meet the observational data at different galactic longitudes (Vallée 2017e).

Also, they employed a 'long bar' (9 kpc diameter) in the Galactic Nucleus, which would strongly perturb the origin of each of the 4 spiral arms (Vallée 2016b – his Section 3.2); no arm perturbation is cited nor explained.

Their segmented arm model is based on radio masers alone, and it does not include any other arm tracers (gas, dust, HI, HII, etc). The subjective addition of five ad hoc "kinks" (l<180°) and four ad hoc "constrained tangencies" (l>180°) in the spiral arms is reminiscent of the subjective addition of a dozen ad hoc "epicycles" in the old but flawed galactic model centered on the Earth (Vallée 2005 – his Fig. 2). Most of these ad hoc 'kinks' and 'constrained tangencies' are not explained physically, and they use no tidal stress from a very nearby galaxy.

**5.2 The Hou & Han (2014) arm model, as fitted to 2500 HII regions**

Their model (in their Fig. 10b and Fig. 13a) places the Cygnus arm at 14.9 kpc along the Galactic Meridian l=180°, or 6.6 kpc beyond the Sun (with $R_{sun}$= 8.3 kpc); see top of Figure 1 here. Their method does not use a global arm pitch angle, and it is limited to fit each arm alone (arm segments, with varying results in their Table 1), with the ensuing large error bars.

Their model is sufficiently close to our's and it will not alter our conclusions: all masers that we put in a new interarm island (Fig. 1) would still be there, and their Cygnus arm would not go through the interarm island (being above it). In their model, the three masers close to the Cygnus arm would still be close, one maser would have just entered the Cygnus arm (G182), but not the other two (G135; G097).

The Cygnus arm crossing the Galactic Meridian is also seen in Fig. 17 in Molina Lera et al (2019), which refers to the same 4-arm logarithmic arm model (Hou et Han 2014, but now with $R_{sun}$=8.5 kpc).

### 5.3 The Koo et al (2017) arm model, as fitted to dense HI gas

The arm model of Koo et al (2017 – their Fig. 5a) fits a single tracer, namely atomic hydrogen, using a flat rotation curve model to get distances. It is not based on starforming regions, per se. They found several 'long arcs' (along main spiral arms) and numerous 'interarm features'. They noted in their Section 3.2 that for the Cygnus arm, there is currently considerable confusion in the identification of their associated HI features.

Thus their Cygnus arm model is 6.2 kpc beyond the Sun, as fitted to dense HI gas, with the $R_{sun}$= 8.34 kpc – see top of Figure 1 here. Again here, their model is sufficiently close to our's and it will not alter our conclusions: all masers in a new interarm island (Fig. 1) would be there, and their Cygnus arm would not go through the interarm island (being above it). The Koo et al (2017) model passes beyond the new interarm island suggested in Fig.1.

Their kinematical model employs velocity-based distances and does not give accurate trigonometric-like distances; kinematic distance errors can be large near the Perseus arm - thus W3(OH) has a parallax distance of 2.0 kpc and a kinematic distance of 4.3 kpc (Xu et al 2006).

### 5.4 Arm model, constructed through a backward integration of stellar orbits, using optical stars

Young open star clusters (of less than 50 million years old) at optical wavelengths, and 'backward orbital integration' of stars, have been employed to determine their birthplaces (Amaral & Lépine, 1997; Dias & Lépine, 2005; Junqueira et al 2015; Dias et al 2019). This method requires a good determination of the stellar distance and intrinsic speed, otherwise the 'backward integration' will be incomplete. This precise distance requirement is only possible observationally near the Sun, owing to instrumental errors, and a refined dust calibration law varying with distance and with galactic longitude. Also, the value of the age of the star cluster must be obtained with a high precision (Fig. 5 in Amaral & Lépine 1997). In all, this requires a good deal of modeling (cluster membership, isochrone fit, dust extinction correction, peculiar speed, narrow dip in rotation curve, arm polynomial/logarithmic fit, etc) and concomittent assumptions – see Dias et al (2019). Basic assumptions are made: all or most star clusters are created in spiral arms, their galactic orbits can be calculated by the derived galactic potential and their present-day accurate space velocity components, etc (Dias & Lépine, 2005).

Some objections can be pointed out here: long integration time, star creation in an interarm or spur, and the Local Arm not being a long density-wave arm. At optical wavelengths, there are almost no open star cluster younger than 5 Myrs (Dias & Lépine 2005), hence the backward integration time must use older star clusters that may have diverged from the pure circular orbit, thus increasing the final position errors. Radio masers are much younger (<1 Myr). A typical lifetime of an O5.3 III star is only 3 Myr (Weidner & Vink 2010). Using longer timescale only adds to the error bars when integrating backward along an orbit. Their use of young open star clusters employs stars already above 10 Myrs, up to 50 Myrs, and thus quite scattered away from their birthplaces (see Fig. 2 in Dias & Lépine 2005 for stellar ages above 30 Myrs). The critical value for the young cluster age is model dependent; different models can be found to get the cluster ages, such as the adopted theoretical isochrones fitting to the photometric data, the adopted reddening, and the adopted spectral type of the stars.

In addition, many nearby stars are being formed in a local spur/bridge or interarm island (not in a long arm) and they would be included in their analysis (interarm stars should have been excluded), and together they could predict an arm quite displaced from its true position.

In their derivation, the Local Arm is taken as a true long arm, yet this is known not to be the case, just a small localized region being affected by an incoming mass from the halo, and/or by the 4:1 Inner Lindblad Resonance (Vallée 2018b). This would affect the gravitational potential in the density wave.

Gaia DR2 optical stars yield distance estimates through trigonometric measurements. Here the majority of optical stars have a large error in their trigonometric precession values, with (f) defined as the relative Gaussian error value over the parallax value often being above 20% (Bailer-Jones 2015), while the error in the inverse of the parallax is not Gaussian. This implies a sufficient bias in the distance determination, for which a suitable probabilistic 'prior' must be applied to give a probability that the source is lying at the far distance. The added priors and assumptions render their arm model more complex, and their result more ambiguous.

The *major* difference in the stellar age (<1 Myr for radio masers, <50 Myrs for optical open star clusters), the use (or not) of the Local Arm (an interarm island, not a long arm), the lack (or not) of dust extinction at the chosen wavelengths, and the simplicity (or multiplicity) of the models employed, together may account for the different results in determining galactic orbits of spiral arms, and where is the true co-rotation of gas and stars versus the spiral pattern.

Their backward-integration model does not predict the location of the Cygnus arm, along the Galactic Meridian.

*Co-rotation*. Backward integration allows one to compute where the spiral pattern was in the past (age = 0), so the difference in space over this time interval can yield the spiral pattern's angular rotation rate. This allows the prediction of a co-rotation radius (when the predicted rotating density wave spiral pattern speed equals the observed orbital rotation speed of the gas and stars). They find a co-rotation value to be near the Sun's circular orbit, namely at a galactic radius of 9.0 kpc (Amaral & Lépine 1997), 8.5 kpc (Dias & Lépine 2005), 8.7 kpc (Junqueira et al 2015), and 8.5 kpc (Dias et al 2019).

While a complex backward integration of optical young open clusters yield results near 8.5 kpc, many radio data yield co-rotation nearer 12 kpc or longer. The use of very young radio masers (age < 1 Myr) located in front of spiral arms in Perseus and Cygnus (but not on the outer arm side) suggests a larger co-rotation radius, as masers are not seen beyond the Cygnus outer arm side (Vallée 2018a; Vallée 2019); hence co-rotation must be beyond a galactic radius of 15 kpc (from Fig. 1 here). Elsewhere, the measured terminal HI gas velocity at radio wavelengths was employed earlier to get a co-rotation radius near 14 kpc (Foster & Cooper 2010 – their Section 3.2.2 and Fig. 4). For the Perseus arm, the location of W3(OH) is on the nearby arm edge (not the outer arm edge), in front of many dust clouds, with the larger CO cloud observed behind (Xu et al 2006); this particular distance ordering (maser and dust to cold gas cloud) indicates that the Perseus arm is located inside the co-rotation radius. Also, Sakai et al (2015 – their Section 4.2) employed VLBI astrometry to find the co-rotation radius near 12.4 kpc. These radio methods to determine orbits and co-rotation are more direct than the optical methods.

### 5.5 Arm model, as fitted to the broad CO 1-0 molecules seen at the arm tangents

For complementarity, we list here our arm model, as fitted to the arm tangents where the CO 1-0 molecule shows a peak intensity when scanning along increasing galactic longitudes. It also corresponds to the 'minimum potential' of the density wave theory (Roberts 1975 - his Fig.2). In our model (Vallée 2017b; Vallée 2017c), the Cygnus arm is at 7.2 kpc beyond the Sun, at $R_{sun}$=8.1 kpc, as fitted to the arm tangents observed in CO 1-0 gas – see top of Figure 1 here.

Age sequence. The spacing of the Cygnus arm, along the Galactic Meridian, shows an offset for each tracer. As measured beyond the Sun's orbit, along the Galactic Meridian, the maser-fit for the Cygnus arm is distant by 4.8 kpc for $R_{sun}$=8.15 kpc (4.9 kpc fo $R_{sun}$ = 8.34 – Section 5.1), while the dense HI gas fit is at 6.1 kpc at $R_{sun}$= 8.15 kpc (6.2 kpc at $R_{sun}$ = 8.34 kpc – Section 5.3), the HII-region-fit is at 6.5 kpc at $R_{un}$=8.15 kpc (6.6 kpc at $R_{sun}$=8.34 kpc – Section 5.2), and the the broad peak CO 1-0 gas fit is at 7.2 kpc at $R_{sun}$=8.15 kpc (Section 5.5). These slightly different arm tracer offsets are cooperating to give the best Cygnus arm location – they are not in disagreement, but show an age sequence. This age sequence of arm tracer offsets (from masers to HI, to HII, and to cold CO gas) altogether reflects the same known sequence, as previously noticed for different starforming tracers looking at the arm tangents (Fig. 4 in Vallée 2017d; Fig.1 in Vallée 2014) and predicted before (Roberts, 1975; Shu 2016). These different locations for the Cygnus arm, seen in different arm tracers, are shown at the top of Figure 1 here.

## 6. Conclusion

We assembled distance and velocity maser data with a known trigonometric parallax (Table 1), excluding those with a large fractional parallax error. Next, we tweaked the CO-fitted 4-arm global model of Vallée (2017b; 2017c). This global 4-arm model employs the arm tangents on both sides of the Galactic Meridian (Galactic Quadrants I and IV), for each arm (Section 2).

We distinguished the trigonometric masers between the Perseus and Cygnus arms, in two groups: those in an interarm island located between the two long arms (similar to the interarm island near the Sun), and those much closer to the Cygnus arm (Figure 1). This new 'interarm island' resembles the 'Local Orion arm' between the Perseus and Sagittarius arms, and the old 'Loop' between the Sagittarius and Scutum arms (Table 2).

We also show the locations of these trigonometric masers in velocity space (Figures 2 and 3), along with the location of the arms and of the interarm islands (Section 3). Other masers and young stars are found inward of the Cygnus arm model (Figure 4); this offset (maser locations versus CO 1-0 based 4-arm model) is interpreted within the density-wave theory (Section 4).

We also discuss whether the observations favor a superposition of a global arm model with interarm islands being inserted (Section 4), or the reconstruction and assembling of several short fragments of long arms (Section 5).

In some arm models (Sections 5.1, 5.2, 5.3, 5.5), the location of the Cygnus arm is quite similar, despite the different approaches at radio wavelengths. Near the Cygnus arm along the Galactic Meridian, these other models (masers, HI, HII, CO 1-0) support the presence of an interarm island concept there. In another arm model (open star cluster; Section 5.4), the results differ substantially, owing to the level of complexities at optical wavelengths.

## Acknowledgements

The figure production made use of the PGPLOT software at NRC Canada in Victoria. I thank an anonymous referee for useful, careful, and historical suggestions.

**Table 1.** Trigonometric masers near the Cygnus (Outer Norma) spiral arm (red curve).

| Name | Gal. Long. (deg.) | Gal. Lat. (deg.) | Distance D [range] (kpc) | Half range (kpc) | Fract. parall. error | Syst. $V_{lsr}$ (km/s) | Reference |
|---|---|---|---|---|---|---|---|
| (1) | (2) | (3) | (4) | (5) | (6) | (7) | (8) |
| G075.29+01.32 | 075.3 | +1.3 | 9.26 [8.85-9.71] | ±0.43 | 0.05 | -58 ±5 | Reid et al (2014); Sanna et al (2012) |
| G090.92+01.48 | 090.9 | +1.5 | 5.85 [4.95-7.14] | ±1.09 | 0.18 | - | Quiroga-Nunez et al (2019) |
| G097.53+03.18 | 097.5 | +3.2 | 7.52 [6.66-8.62] | ±0.08 | 0.13 | -73 ±5 | Reid et al (2014); Hachisuka et al (2015) |
| G135.27+02.79 | 135.3 | +2.8 | 5.99 [5.62-6.41] | ±0.40 | 0.07 | -72 ±3 | Reid et al (2014); Hachisuka et al (2009) |
| G160.14+03.16 | 160.1 | +3.2 | 4.10 [4.00-4.20] | ±0.10 | 0.02 | -18 ±5 | Reid et al (2014) |
| G168.06+00.82 | 168.1 | +0.8 | 4.98 [4.44-5.65] | ±0.60 | 0.12 | -28 ±5 | Hachisuka et al (2015) |
| G182.67-03.26 | 182.7 | -3.3 | 6.71 [6.25-7.24] | ±0.50 | 0.07 | -07 ±10 | Reid et al (2014); Hachisuka et al (2015) |
| G196.45-01.68 | 196.4 | -1.7 | 4.15 [3.95-4.33] | ±0.19 | 0.05 | +18 ±2 | Quiroga-Nunez et al (2019) |
| G217.80+01.05 | 217.8 | +1.1 | 6.14 [5.59-6.80] | ±0.19 | 0.10 | +71 ±5 | Quiroga-Nunez et al (2019); Sparks et al 2008) |

Notes :
In column 1, taking all trigonometric maser sources, identified as located in the 'Cygnus' (outer Norma) arm in Reid et al (2014) or Hachisuka et al (2015), or Quiroga-Nunez et al (2019). Other names are used: S269 (G196.45-01.68) and V838 (G217.80+01.05).

In column 6, we excluded sources with a large fractional parallax error f > 0.2. All distances are trigonometric; the published parallax (p, in mas) was converted to a distance (D, in kpc) through the equation D = 1/p.

In column 8, when two or more references are given, the data from the first one were adopted. Thus the Reid et al (2014) and Quiroga-Nunez et al (2019) system was employed, having a solar motion of (U,V,W) =(10.5, 14.4, 8.9) km/s, which is very similar to the Hachisuka et al (2015) system of (11.1, 12.2, 7.3) km/s.

**Table 2.** **Known interarm Islands, in the disk of the Milky Way**

| Island name | Between which arms | Gal. longit. (o) | Radial vel. (km/s) | Distance (kpc) | Width[1] (kpc) | Reference |
|---|---|---|---|---|---|---|
| (1) | (2) | (3) | (4) | (5) | (6) | (7) |
| Cygnus island | Perseus & Cygnus | 160° to 220° | -30 to +40 | 5 | 6x2 | Fig. 1 here |
| Armlet, ring | Sagittarius & Perseus | 0° to 360° | -20 to + 20 | < 2 | 4x2 | Vallée (2018b) |
| Potential loop | Scutum & Sagittarius | 32° to 39° | +65 to + 90 | 4.4 | 2x0.5 | Rigby et al (2016) |

Notes:
   In column (5), the distance scale of 8.15 kpc was taken for the distance of the Sun to the Galactic Center.
   In column (6), the width is given as: the long axis (usually in galactic longitude) by the short axis.

**Table 3.** Maximum distance of the observed logarithmic spiral arms, in nearby disc galaxies, as seen at a wavelength near 21 cm HI or continuum (if equal or longer than 12 Mpc)

| Disc Galaxy Messier no. | Distance of Galaxy (Mpc; reference) | Logarithmic spiral arm out to a radial distance (arc min) (reference) | Corresponding galactic radius (kpc) |
|---|---|---|---|
| M 74  | 10.0 | 9' (Fig. 3 in Shostak & van der Kruit 1984) | 26 |
| M101  | 7.2  | 12 (Fig. 9 in Allen & Goss 1979)            | 25 |
| M 81  | 3.25 | 24 (Fig.8 in Rots & Shane 1975)             | 23 |
| M 106 | 6.6  | 9.8 (Fig. 1 in van Albada & Shane 1975)     | 19 |
| M 31  | 0.8  | 80 (Fig. 6 in Braun 1991)                   | 19 |
| M 51  | 9.7  | 4.2 (Fig. 4 in Segalovitz 1977)             | 12 |

# Figure Captions

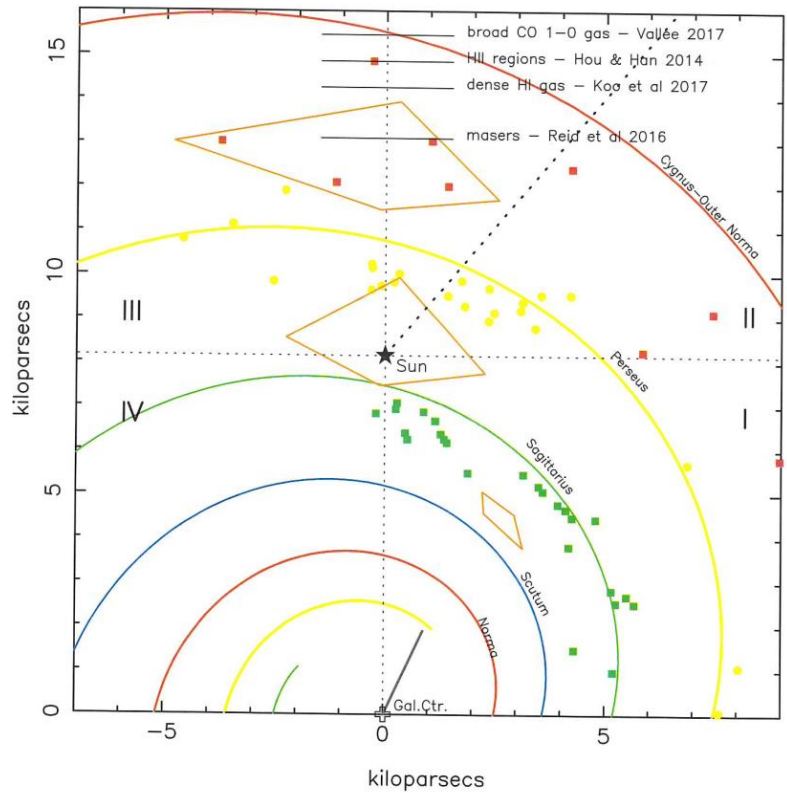

**Figure 1.** A face-on plot of the disk of the Milky Way. The Sun is at (0 kpc, 8.15 kpc) and the Galactic Center is at (0, 0). Locations of the known interarm islands are shown (orange curves), between the Scutum and Sagittarius arms, between the Sagittarius and Perseus arm, and between the Perseus and Cygnus arms. Squares represent the positions of trigonometric masers in the galactic disk. Galactic Quadrants I to IV are shown, as well as the 4 spiral arms. Near the top of the Figure, we show he locations of the Cygnus arm on the Galactic Meridian, from other models using other arm tracers. A similar figure, without the interarm islands, appeared in Vallée (2017c - his fig. 5a).

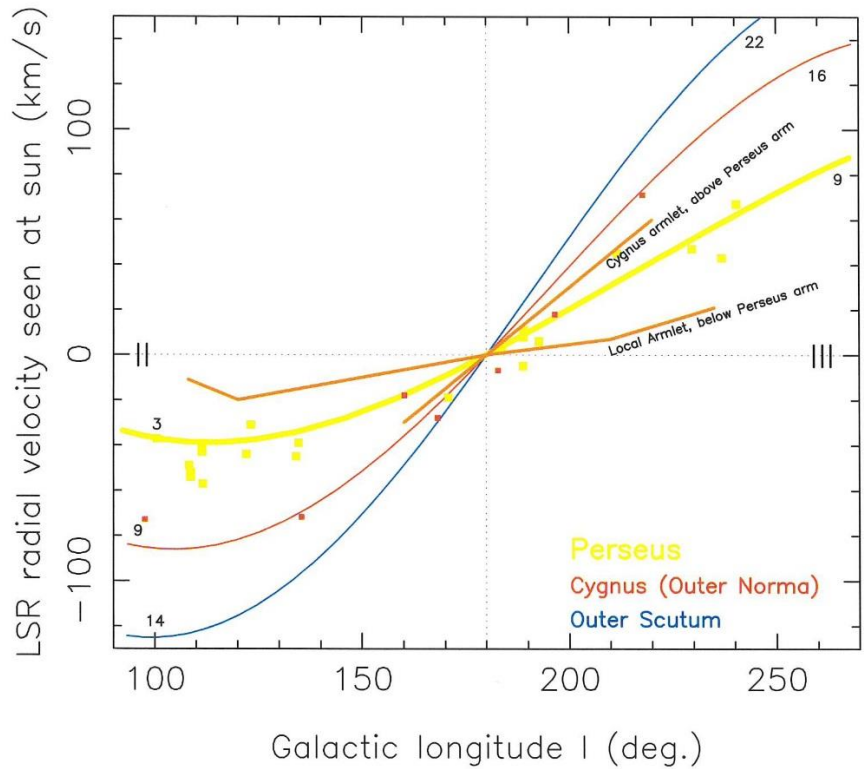

**Figure 2.** Radial velocity space (vertical axis) and galactic longitudes (horizontal axis) are shown, looking away from the Galactic Center. Locations of two interarm islands are shown (orange lines). The spiral arms are shown as continuous curves. Numbers on the arms indicate the rough distance of that arm point to the Sun. Trigonometric masers (squares) are shown for Cygnus (red) and Perseus (yellow) arm. A similar figure, without the interarm islands, appeared in Vallée 2017c (his Fig. 5b). This model has been upgraded for the refined values of $R_{sun}$=8.15 kpc aad the LSR orbital velocity $V_{lsr}$=233.3 km/s.

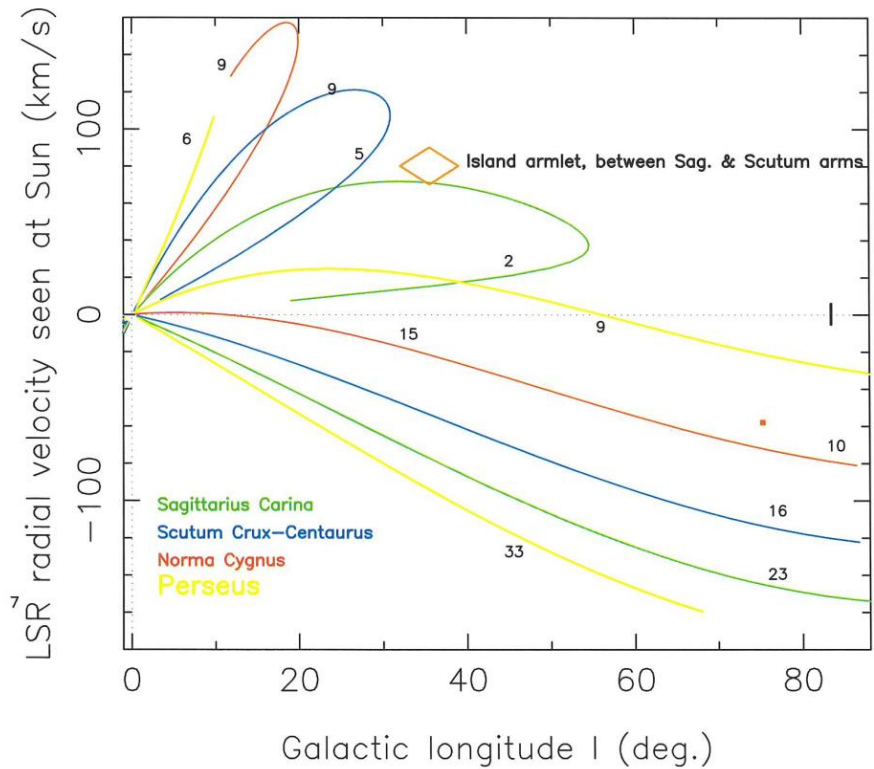

**Figure 3.** Plot of radial velocity (km/s) versus galactic longitude (°), looking toward the inner Galaxy. The location of one interarm island is shown (orange curve), near l=35°). The red square at l=75° is a maser near the Cygnus arm, at a solar distane near 9.3 kpc (from Table 1). The spiral arms are shown (continuous curves), with numbers indicating a rough distance of that point to the Sun. A similar figure (without the island) appeared In fig.4 in Vallée (2017b). This model uses the latest $R_{sun}$=8.15 kpc and $V_{lsr}$=233.3 km/s.

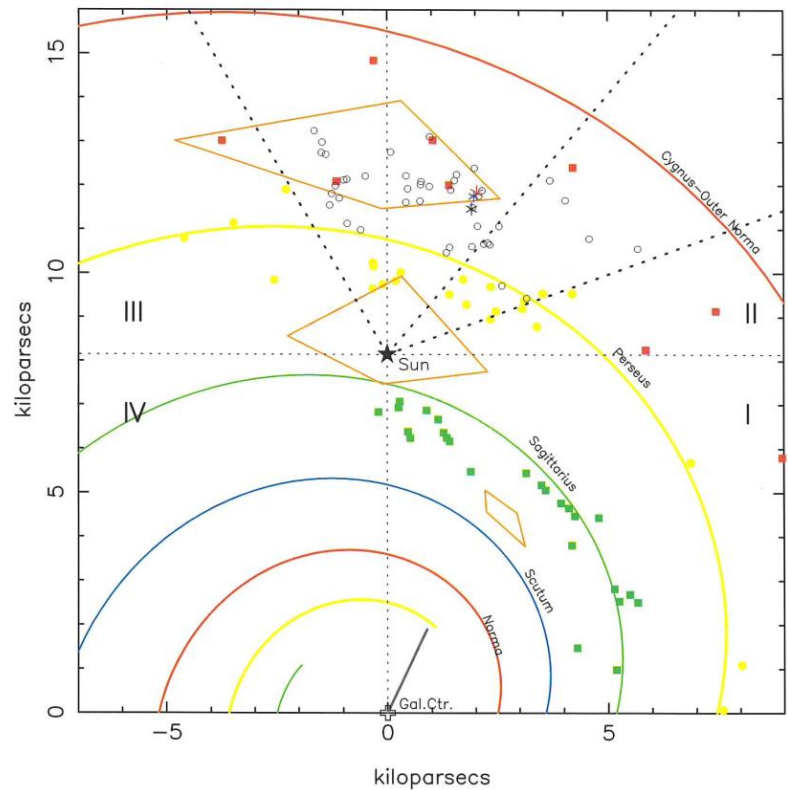

**Figure 4.** This plot, similar to Figure 1, now shows the addition of the young open star cluster Waterloo 1 (black asterisk, near l=150°), Sh-207 near l=151° (red asterisk) and Sh 208 near l=151.3° (blue asterisk) as well as numerous HII regions (small open circles) between l=110° and 210° between the Perseus arm and the Cygnus arm. Galactic longitudes 110°, 140°, and 210° are shown by short dashes lines (see Sections 3.1 and 3.3).